\documentclass[%
 reprint,
superscriptaddress,
nofootinbib,
 amsmath,amssymb,
 aps,
]{revtex4-2}

\usepackage[dvipdfmx]{graphicx}
\usepackage{dcolumn}
\usepackage{bm}
\usepackage{bbm}
\usepackage[colorlinks=true, allcolors=blue]{hyperref}
\usepackage[usenames]{color}

\newcommand\barparena[1]{\overset{%
   \scriptscriptstyle(-)}{#1}}

\newcommand{\dd}{\mathrm{d}}
\newcommand{\ii}{\mathrm{i}}

\newcommand{\GF}{G_{\mathrm{F}}}
\newcommand{\abs}[1]{\left\lvert#1\right\rvert}

\begin{document}

\title{Characterizing quasi-steady states of fast neutrino-flavor conversion by stability and conservation laws}

\author{Masamichi Zaizen}
\email{zaizen@heap.phys.waseda.ac.jp}
\affiliation{Faculty of Science and Engineering, Waseda University, Tokyo 169-8555, Japan}

\author{Hiroki Nagakura}
\affiliation{Division of Science, National Astronomical Observatory of Japan, 2-21-1 Osawa, Mitaka, Tokyo 181-8588, Japan}

\date{\today}

\begin{abstract}
The question of what ingredients characterize the quasi-steady state of fast neutrino-flavor conversion (FFC) is one of the long-standing riddles in neutrino oscillation.
Addressing this issue is necessary for accurate modeling of neutrino transport in core-collapse supernova and binary neutron star merger.
Recent numerical simulations of FFC have shown, however, that the quasi-steady state is sensitively dependent on boundary conditions in space, and the physical reason for the dependence is not clear at present.
In this study, we provide a physical interpretation of this issue based on arguments with stability and conservation laws.
The stability can be determined by the disappearance of ELN(electron neutrino-lepton number)-XLN(heavy-leptonic one) angular crossings, and we also highlight two conserved quantities characterizing the quasi-steady state of FFC:
(1) lepton number conservation along each neutrino trajectory and (2) conservation law associated with angular moments, depending on boundary conditions, for each flavor of neutrinos.
We present an analytic prescription that matches the results of the nonlinear simulations presented in this work.
This study represents a major step forward a unified picture determining asymptotic states of FFCs.
\end{abstract}

\maketitle


\section{Introduction}
Neutrinos are copiously produced in the dense core of compact objects such as core-collapse supernovae (CCSNe) and binary neutron star mergers (BNSMs).
They have key roles on the astrophysical environments through weak interactions with matters.
Proper descriptions of the neutrino transports, matter interactions, and flavor conversions, are crucial to understand the physical processes in these astrophysical phenomena.

Flavors of neutrinos can change during their propagation and affect the dynamics and nucleosynthesis in CCSNe and BNSMs.
In dense neutrino media, their self-interactions are potentially dominant, leading to collective neutrino oscillations \cite{Pantaleone:1992,Pantaleone:1992a,Sigl:1993,Duan:2010,Mirizzi:2016,Chakraborty:2016}.
In particular, fast neutrino-flavor conversion (FFC), which is dictated only by the self-interactions, is prominent on scales by $\sim (\GF n_{\nu})^{-1}$ \cite{Sawyer:2016,Chakraborty:2016a,Richers:2022b}.
The necessary and sufficient conditions for FFC are equivalent to the presence of angular crossings in $(f_{\nu_e}-f_{\bar{\nu}_e})-(f_{\nu_X}-f_{\bar{\nu}_X})$ (hereafter called ELN-XLN angular crossings), which is in line with linear stability analysis \cite{Izaguirre:2017,Morinaga:2022,Dasgupta:2022}.
Occurrences of ELN (electron neutrino-lepton number) crossings have been surveyed for realistic models on CCSNe and BNSMs \cite{Dasgupta:2018,Abbar:2019,Abbar:2020,Abbar:2020b,Abbar:2021,Abbar:2023,DelfanAzari:2020,Glas:2020,Morinaga:2020,Johns:2021,Nagakura:2019,Nagakura:2021,Nagakura:2021b,Nagakura:2021c,Harada:2022}.
Very recently, it has been suggested that FFC can substantially alter the energy flux of electron-type neutrinos, which reduces the neutrino heating in the gain region of CCSN core \cite{Nagakura:2023,Ehring:2023}.
This exhibits that the accurate modeling of FFC is necessary to understand the explosion mechanism of CCSN.

Recently, the asymptotic states of FFC have attracted significant attention.
In a homogeneous and collisionless system, FFC occurs quasi-periodically, which is an analogy to the pendulum motion \cite{Johns:2020}.
On the other hand, if advection (or inhomogeneity) and collisions are included in FFCs, the periodicity disappears;
instead, the system evolves toward a quasi-steady (or asymptotic) state.
It has been suggested that collisions with matter affect the flavor evolution through collisional decoherence and bring out its damping or enhancement on the flavor contents \cite{Martin:2021,Shalgar:2021a,Hansen:2022,Kato:2021,Kato:2022,Kato:2023,Sasaki:2022a,Sigl:2022,Johns:2022,Johns:2021a,Johns:2022a,Lin:2022a,Xiong:2023a,Xiong:2022b,Liu:2023}.
Neutrino advection also triggers the flavor coherence cascade into smaller scales, and the entire system eventually reaches a nonlinear saturation \cite{Richers:2022,Bhattacharyya:2020,Bhattacharyya:2021,Bhattacharyya:2022,Wu:2021,Richers:2021a,Zaizen:2021a,Zaizen:2023a,Nagakura:2022,Nagakura:2022a,Nagakura:2023a}.
One notable feature appearing in the nonlinear saturation is that a {\it flavor equipartition} on one side of ELN angular distributions has been commonly observed.
Understanding such a generic feature in the nonlinear phase is key information to incorporate the effects of FFC into global radiation-hydrodynamic simulations of CCSNe \cite{Ehring:2023} and BNSMs \cite{Just:2022,Fernandez:2022} by phenomenological approaches.

Growing attention has also been paid to develop analytical models to better incorporate the angular structure of the FFC \cite{Bhattacharyya:2022,Zaizen:2023a,Nagakura:2023a}.
In a series of our studies \cite{Nagakura:2022,Nagakura:2022a,Nagakura:2023a,Zaizen:2023a}, we suggested that there are hints in ELN-XLN angular distributions to characterize the quasi-steady state of FFC.
For instance, the disappearance of ELN-XLN angular crossing is a generic feature.
This argument is in line with the stability analysis.
In Ref.\,\cite{Zaizen:2023a}, we provided a simple and analytic scheme to determine a neutrino distribution in a quasi-steady state under the periodic boundary condition, which is in agreement with the results of numerical simulations.

On the other hand, it has been reported that different quasi-steady state is achieved under the Dirichlet boundary condition \cite{Nagakura:2022,Nagakura:2022a,Nagakura:2023a}.
The disappearance of ELN-XLN angular crossings is commonly observed even in the Dirichlet case, but angular directions in which a flavor equipartition occurs are qualitatively different from in the periodic one, even if the initial neutrino distributions are set to be the same.
It indicates that the frameworks driving the asymptotic behaviors vary due to the boundary conditions, but the dependence remains obscure.
It should also be mentioned that the Dirichlet boundary condition, from which neutrinos with pure flavor states are injected into the simulation box, would represent the case in the transition layer from no flavor conversion to FFC.
Understanding the physical reason for how the boundary condition affects the quasi-steady state of FFCs motivates this study.
We also generalize our analytic scheme to be applied in both boundary conditions.

This paper is organized as follows.
In Sec.\,\ref{Sec.II:quasi}, we argue the quasi-steady state of FFC analytically, and then explain why and how the quasi-steady state of FFC depends on the boundary condition.
Based on the consideration, we provide a simple analytic scheme, which corresponds to a generalization of our previous one in \cite{Zaizen:2023a,Nagakura:2023a}.
In Sec.\,\ref{Sec.III:NumSim}, we perform one-dimensional FFC simulations to strengthen our arguments.
Finally, we summarize our conclusions in Sec.\,\ref{sec:conclusion}.

\section{Analytic description of quasi-steady state of FFC}\label{Sec.II:quasi}
\subsection{Basic equation}\label{subsec:basiceq}
Neutrino flavor conversion can be described by a quantum kinetic equation (QKE) for neutrino density matrix $\rho$ \cite{Sigl:1993}:
\begin{align}
    \left(\partial_t+\boldsymbol{v}\cdot\nabla\right)\rho = -\ii\left[\mathcal{H},\,\rho\right] + \mathcal{C},
    \label{eq:QKE}
\end{align}
where $\mathcal{C}$ denotes the collisions with matters, and $\mathcal{H}$ represents the Hamiltonian of neutrino oscillations including vacuum, matter, and neutrino self-interaction term.
The Hamiltonian is
\begin{align}
    \mathcal{H} = U\frac{\Delta m^2}{2E}U^{\dagger} + \sqrt{2}\GF\, v^{\mu}\Lambda_{\mu} + \sqrt{2}\GF\int\dd\Gamma^{\prime}\,v^{\mu}v_{\mu}^{\prime}\,\rho^{\prime},
\end{align}
where $\Gamma$ specifies the neutrino energy $E$ and the velocity direction $\boldsymbol{v}$, and the phase space integration is $\int\dd\Gamma = \int^{\infty}_{0}E^2\dd E\int\dd\boldsymbol{v}/(2\pi)^3$ in the flavor-isospin convention where $\bar{\rho}(E)\equiv -\rho(-E)$.
Here, in the first term, $U$ is the Pontecorvo-Maki-Nakagawa-Sakata matrix composed of a mixing angle $\theta_{\mathrm{mix}}$ and $\Delta m^2$ is a squared neutrino mass difference.
The second term represents the matter oscillation, where $v^{\mu}=(1,\boldsymbol{v})$ and $\Lambda_{\mu}=\mathrm{diag}[\{j_{\mu}^l\}]$ with being the lepton number current for charged leptons $l$.
And the last term corresponds to neutrino self-interactions inducing collective neutrino oscillation.
Throughout this work, we consider a one-dimensional problem in flat-space time, which guarantees no neutrino advection in momentum space, and the collision term is neglected just for simplicity.
It is worth to note that our arguments can be applied to more general cases by relaxing these simplifications.
The study is currently underway, and the result will be published in the forthcoming paper.

\subsection{Stability and conservation laws}
\label{subsec:characterizing}
The stability of flavor conversion in the quasi-steady state after the nonlinear saturation can be evaluated on the analogy from linear analysis on the off-diagonal term of the neutrino density matrix \cite{Izaguirre:2017,Morinaga:2022,Dasgupta:2022}.
In the nonlinear regime, we need to consider the mode couplings between the diagonal and the off-diagonal components, but we here focus on the overall trends.
Then, the stability of FFC can be attributed to the disappearance of ELN-XLN angular crossings in the spatially- or temporal-averaged domain \cite{Nagakura:2022a,Zaizen:2023a}.
FFC works to eliminate the angular crossings in which flavor equipartition (equilibrium) occurs in some angular regions \cite{Bhattacharyya:2020,Bhattacharyya:2021,Bhattacharyya:2022,Wu:2021,Richers:2021a,Richers:2022,Zaizen:2023a}.
Note that this trend is generic regardless of boundary conditions (see Ref.\,\cite{Nagakura:2023}).

Below, we make a statement that two conservation laws are the other key ingredients to characterize the quasi-steady state of FFC, and they also account for the difference between periodic and Dirichlet boundary conditions.
The first conserved quantity is the lepton number along each neutrino trajectory.
This can be seen by taking the trace of Eq.\,\eqref{eq:QKE}.
Since the trace part of the oscillation term in the right-hand side of Eq.\,\eqref{eq:QKE} is zero, the net neutrino distribution, $f_{\nu_e}+f_{\nu_x}$, is conserved along each neutrino trajectory unless the collision term is included.
This indicates that the neutrino distribution in the asymptotic state, $f^a(\boldsymbol{v})$, can be given using a linear combination of initial distributions, $f^0_{\alpha}(\boldsymbol{v})$ for $\alpha$ flavor, by
\begin{align}
    &f^{a}_{\mathrm{ELN}}(\boldsymbol{v}) = P_{\mathrm{ELN}}(\boldsymbol{v})f^0_{\mathrm{ELN}}(\boldsymbol{v}) + \left[1-P_{\mathrm{ELN}}(\boldsymbol{v})\right]f^0_{\mathrm{XLN}}(\boldsymbol{v}) \\
    &f^{a}_{\mathrm{XLN}}(\boldsymbol{v}) = \frac{1}{2}\left[1-P_{\mathrm{ELN}}(\boldsymbol{v})\right]f^0_{\mathrm{ELN}}(\boldsymbol{v}) + \frac{1}{2}\left[1+P_{\mathrm{ELN}}(\boldsymbol{v})\right]f^0_{\mathrm{XLN}}(\boldsymbol{v}),
\end{align}
where $P_{\mathrm{ELN}}$ represents a survival probability of ELN.
In the fast limit, $\Delta m^2\to0$, the survival probability is the same for neutrinos ($P_{ee}$) and anti-neutrinos ($\bar{P}_{ee}$).
In the above expression, we also assume $f^0_{\mathrm{XLN}}=f^0_{\mathrm{MuLN}}=f^0_{\mathrm{TauLN}}$, which is a reasonable condition in CCSNs and BNSMs, within the three-flavor framework.
In the two-flavor framework, $f^{a}_{\mathrm{XLN}}$ is rewritten as
\begin{align}
    f^{a}_{\mathrm{XLN}}(\boldsymbol{v}) = \left[1-P_{\mathrm{ELN}}(\boldsymbol{v})\right]f^0_{\mathrm{ELN}}(\boldsymbol{v}) + P_{\mathrm{ELN}}(\boldsymbol{v})f^0_{\mathrm{XLN}}(\boldsymbol{v}).
\end{align}

The second conserved quantity is associated with the zeroth and first angular moments.
To capture the essence, we assume one-dimensional space in the following discussion, while the generalization to multi-dimensional case is straightforward.
Integrating the QKE over the momentum space of neutrinos, we can obtain the following conservative form of moment equation in the fast limit,
\begin{align}
    \partial_t H_E + \partial_z H_F = 0,
    \label{eq:QKE_conserv}
\end{align}
where $H_E$ and $H_F$ represent number density and flux of neutrinos for each flavor, respectively:
\begin{align}
    &H_E = \sqrt{2}\GF\int\dd\Gamma\, \rho, \\
    &H_F = \sqrt{2}\GF\int\dd\Gamma\, v_z \rho.
\end{align}

In the periodic boundary condition, the flux term is canceled with each other on the boundary surfaces when we integrate Eq.\,\eqref{eq:QKE_conserv} over the real space.
More specifically, it can be written as,
\begin{align}
    \partial_t \langle H_E \rangle = H_F\rvert_{\mathrm{outer}} - H_F\rvert_{\mathrm{inner}} = 0,
    \label{eq:periodSPinteg}
\end{align}
while $\langle \rangle$ represents the spatially integrated quantity.
This provides strong constraints on each neutrino distribution for ELN and XLN during the flavor evolution.
In fact, the lepton number for each flavor of neutrinos in the simulation box is constant in time, even when the system has not yet reached a quasi-steady state.
The lepton number conservation constraints the behaviors of flavor conversion and leads to the asymptotic states with less flavor conversion on the other side of angular crossings under the periodic boundary condition \cite{Bhattacharyya:2020,Bhattacharyya:2021,Bhattacharyya:2022,Wu:2021,Richers:2021a,Duan:2021}.
This is the rationale behind the fact that the ELN-XLN angular crossing disappears in the quasi-steady state while keeping the lepton number of each flavor of neutrinos constant \cite{Zaizen:2023a}.

One thing we do notice here is that the periodic boundary condition does not guarantee that the flux is constant in time.
In general, $H_F$ does not need to be homogeneous, and, more importantly, it can change in time while FFC evolves towards the asymptotic state.
This indicates that $H_F$ in the initial condition is different from that in the quasi-steady state.
On the other hand, when the quasi-steady state achieves, the flux needs to be almost constant in space,
\begin{align}
    \partial_z H_F \sim 0,
    \label{eq:fluxcon-quasi}
\end{align}
otherwise the zeroth angular moment evolves with time (see Eq.\,\eqref{eq:QKE_conserv}).
This condition should be satisfied in any quasi-steady state regardless of boundary conditions.
This argument provides the key to understanding the difference in quasi-steady states between periodic- and Dirichlet boundaries, as discussed below.

Different from periodic boundary condition, neutrino fluxes on the Dirichlet boundaries are not identical between inner and outer ones, $\left.H_F\right|_{\mathrm{inner}}\neq \left.H_F\right|_{\mathrm{outer}}$, indicating that $\langle H_E \rangle$ is no longer constant in time.
This indicates that the quasi-steady state can not be determined from the $\langle H_E \rangle$ conservation.
As discussed above, however, the flux needs to be
spatially constant in any quasi-steady state.
This condition can be directly used in the case of the Dirichlet boundary condition.

To highlight the key point, we consider the following situation.
The neutrino in pure flavor eigenstate (but having an ELN-XLN angular crossing in momentum space) is injected constantly in time from one of the boundaries (which corresponds to a Dirichlet boundary condition), and we also impose a free boundary condition in the other one.
This corresponds to a similar situation in Refs.\,\cite{Nagakura:2022,Nagakura:2022a,Nagakura:2023a}.
In this case, $H_F$ at the boundary with the Dirichlet condition is fixed in time, indicating that it should be the same in the quasi-steady state.
By using Eq.\,\eqref{eq:fluxcon-quasi} that should be satisfied in any quasi-steady states, we can obtain the following condition
\begin{align}
    H_F^a (z) = H_F(t=0,z=0),
\end{align}
where $z=0$ corresponds to the boundary where the Dirichlet condition is imposed, and $H_F^a (z)$ represents the neutrino flux in the quasi-steady state.
This argument leads to an important conclusion.
The quasi-steady state in this situation is characterized by the flux conservation, instead of the number conservation in the periodic case.

\subsection{Analytic determination of quasi-steady state}\label{subsec:modeling}
Following the above argument, we provide an analytic scheme to determine the quasi-steady state of FFC, which corresponds to the extended one from Refs.\,\cite{Zaizen:2023a,Nagakura:2023a}.
It should be mentioned that what we need to do is only to determine angular distribution of $P_{\mathrm{ELN}}$, which is by virtue of the lepton number conservation along each neutrino trajectory.
By using the initial neutrino distributions, we first quantify the negative part of ELN-XLN angular moment $A_m$ and positive one $B_m$ as
\begin{align}
    A_m &\equiv \abs{\int_{v_z^m G^{ex}_{v}<0}\dd\Gamma\, v_z^m G_v^{ex}} \\
    B_m &\equiv \abs{\int_{v_z^m G^{ex}_{v}>0}\dd\Gamma\, v_z^m G_v^{ex}},
    \label{eq:angular_moments}
\end{align}
where $G^{ex}_v$ denotes ELN-XLN angular distribution.
The zeroth moment, $m=0$, is associated with the neutrino-lepton number and the first moment, $m=1$, is for the flux.
It is necessary to eliminate ELN-XLN angular crossings to achieve the stability while satisfying the conservation laws on the number or the flux, depending on boundary conditions.

Because of the constraint of either ELN-XLN number or flux conservation, its sign, which can be estimated in the initial condition, determines the angular directions in which a flavor equipartition is achieved.
For the negative case, $B_m-A_m<0$, the survival probability of electron-type neutrinos can be approximately estimated as the following box-like function:
\begin{align}
    P_{\mathrm{ELN}} = \barparena{P}_{ee} =
    \begin{cases}
        p_{\mathrm{eq}}  &\mathrm{for}\,\, v_z^m G_v^{ex} > 0\\
        1 - (1-p_{\mathrm{eq}})\dfrac{B_m}{A_m} &\mathrm{for}\,\, v_z^m G_v^{ex} < 0,
    \end{cases}
    \label{eq:approx_1}
\end{align}
where $p_{\mathrm{eq}}$ is a survival probability for a flavor equipartition. 
In the angular region with $P_{ee} = p_{\mathrm{eq}}$, the flavor equipartition is achieved, i.e., ELN-XLN being zero.
For the positive case, $B_m-A_m>0$, we obtain
\begin{align}
    P_{\mathrm{ELN}} = \barparena{P}_{ee} =
    \begin{cases}
        p_{\mathrm{eq}}  &\mathrm{for}\,\, v_z^m G_v^{ex} < 0\\
        1 - (1-p_{\mathrm{eq}})\dfrac{A_m}{B_m} &\mathrm{for}\,\, v_z^m G_v^{ex} > 0.
    \end{cases}
    \label{eq:approx_2}
\end{align}
This analytical scheme for $m=0$ becomes identical to our previous one under the periodic boundary condition (see Eqs.\,(22) and (23) in Ref.\,\cite{Zaizen:2023a}).
Meanwhile, the case with $m=1$ represents the extended version of our scheme for the Dirichlet boundary.

It is interesting to note that, if the sign of $B_1-A_1$ is opposite to that of $B_0-A_0$, the angular direction where the flavor equipartition occurs becomes the opposite between periodic and Dirichlet boundaries.
Also, Eq.\,\eqref{eq:approx_1} and Eq.\,\eqref{eq:approx_2} suggest full flavor equilibrium in the entire angular directions for the symmetric flavor case, $B_m=A_m$.
This indicates that the flavor equipartition in the entire angular domain can be achieved even in the Dirichlet boundary condition, which has not been observed in previous studies \cite{Wu:2021}.
In the following section, we provide evidence by numerical simulations that these intriguing features actually emerge in the quasi-steady state, as predicted by our analytic scheme.

\begin{figure*}[t]
    \centering
    \includegraphics[width=0.95\linewidth]{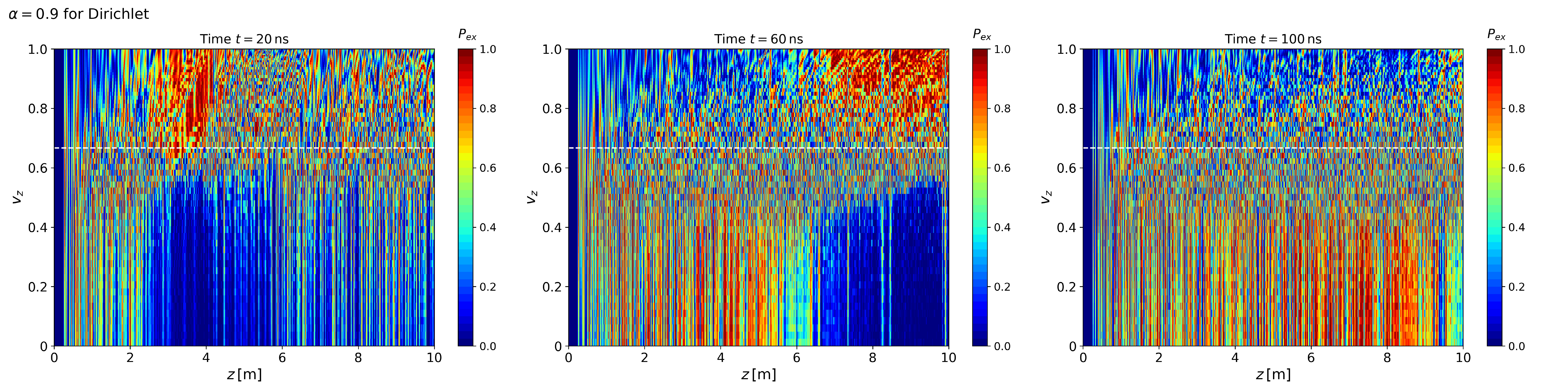}
    \includegraphics[width=0.95\linewidth]{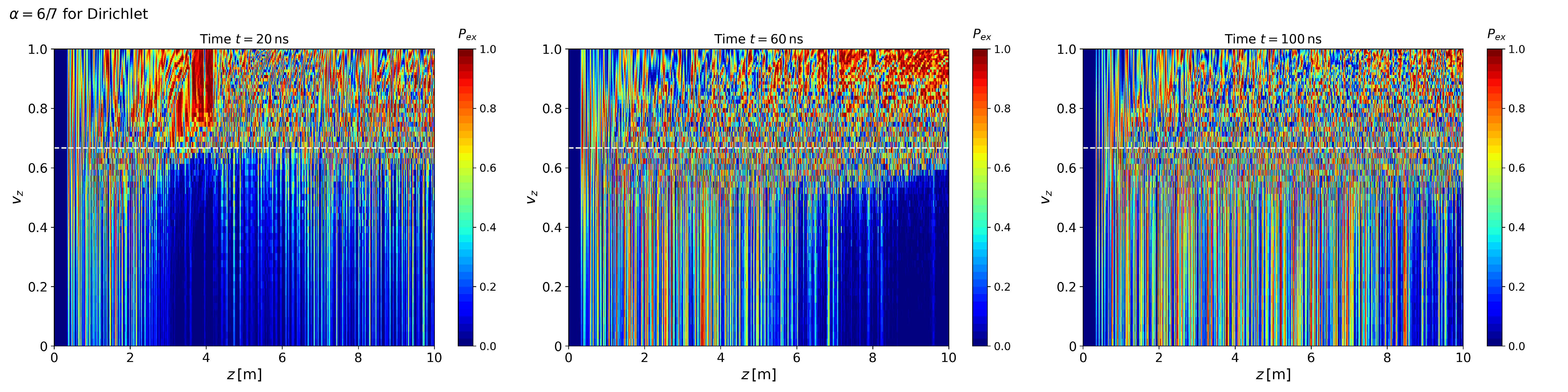}
    \caption{Transition probability $P_{ex}$ for $\alpha=0.9$ (top) and for $\alpha=6/7$ (bottom) at different time snapshots ($t=20, 60$, and $100\mathrm{\,ns}$).
    The white dashed line in each panel draws the location of a crossing in initial ELN angular distribution.}
    \label{fig:t_evo_Pex}
\end{figure*}
\begin{figure}[t]
    \centering
    \includegraphics[width=0.95\linewidth]{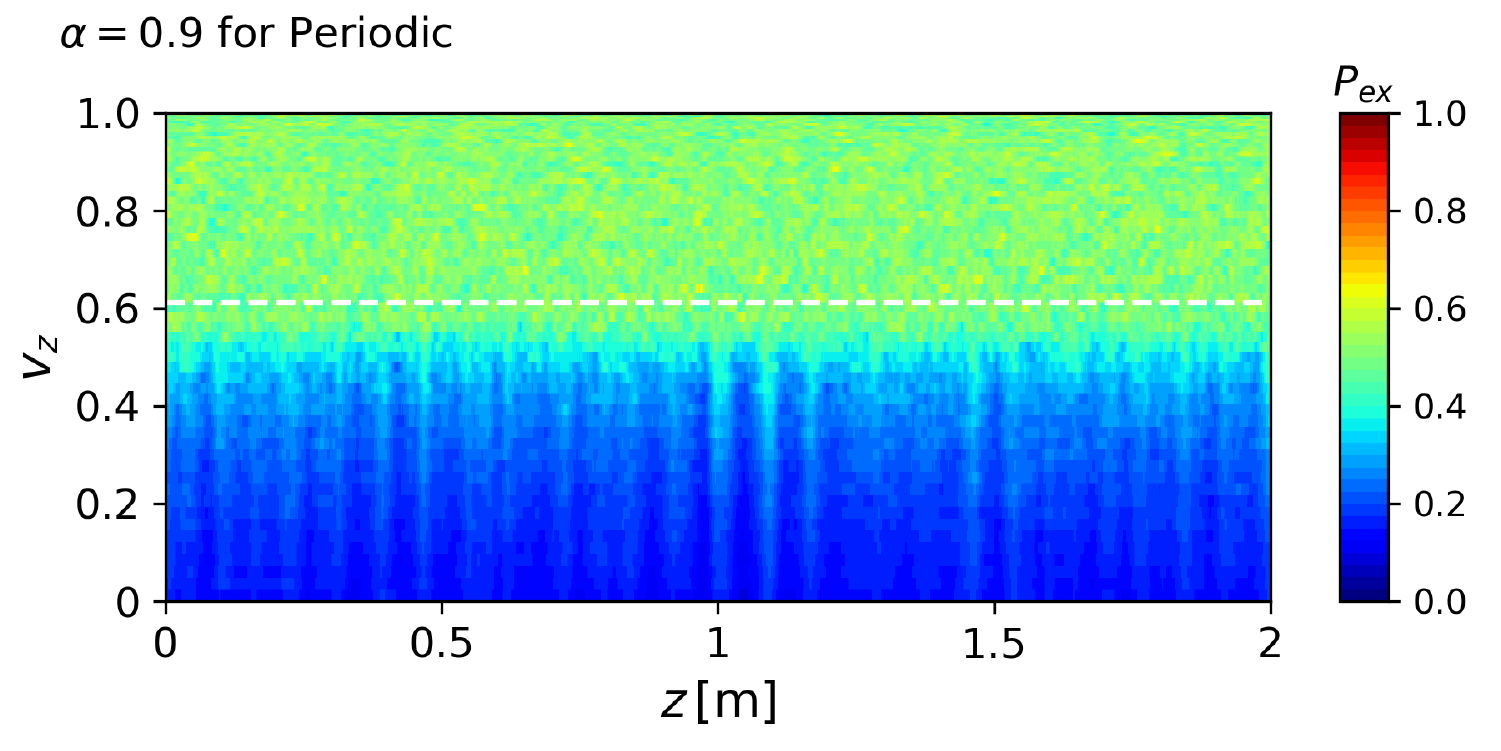}
    \includegraphics[width=0.95\linewidth]{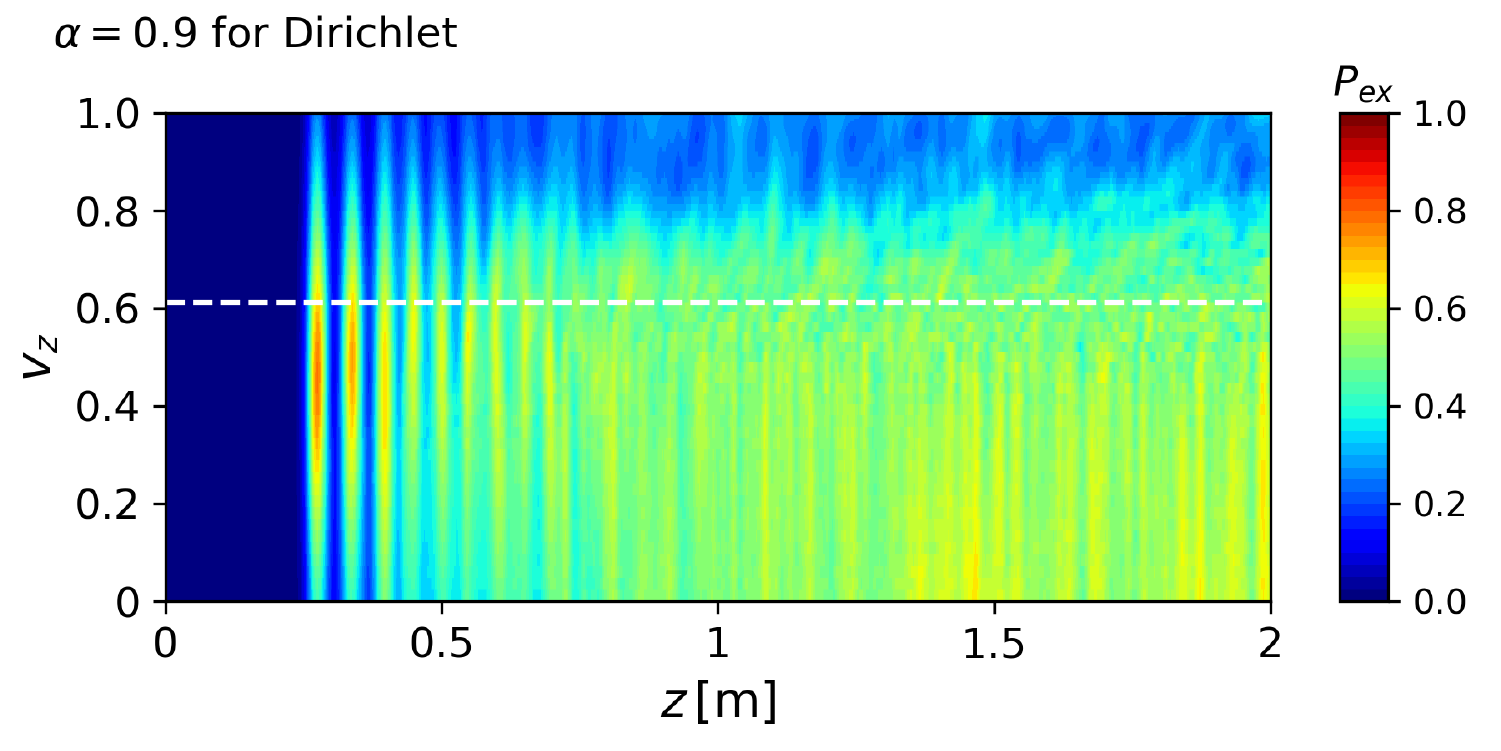}
    \includegraphics[width=0.95\linewidth]{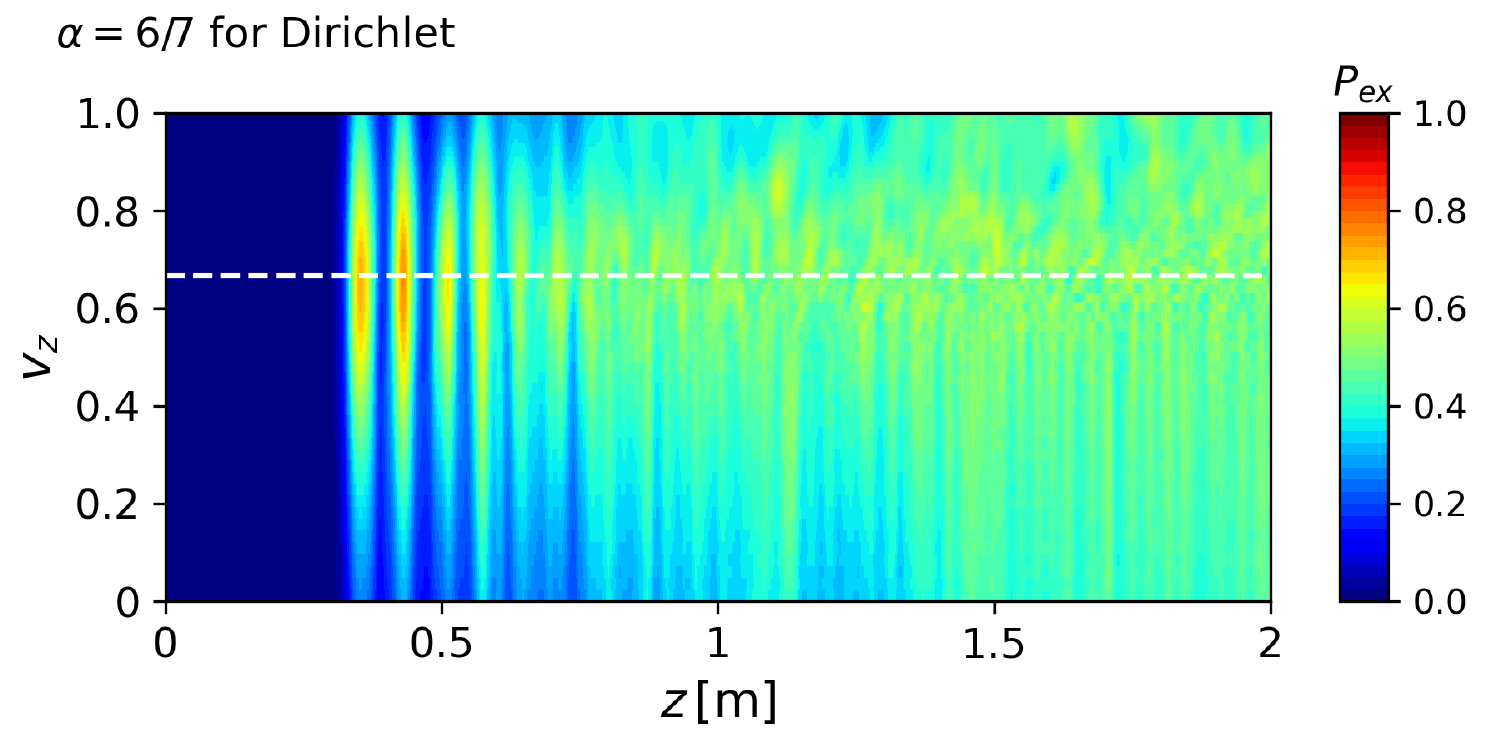}
    \caption{Transition probability $P_{ex}$ averaged in the window of $40\,\mathrm{ns}\leq t \leq 100\,\mathrm{ns}$ during the quasi-steady state for $\alpha=0.9$ (top and middle) and $\alpha=6/7$ (bottom).
    Only top panel is for periodic case and the others for Dirichlet case.
    The white dashed line in each panel draws the location of a crossing in initial ELN angular distribution.}
    \label{fig:t_avg_Pex}
\end{figure}
\begin{figure}[t]
    \centering
    \includegraphics[width=0.95\linewidth]{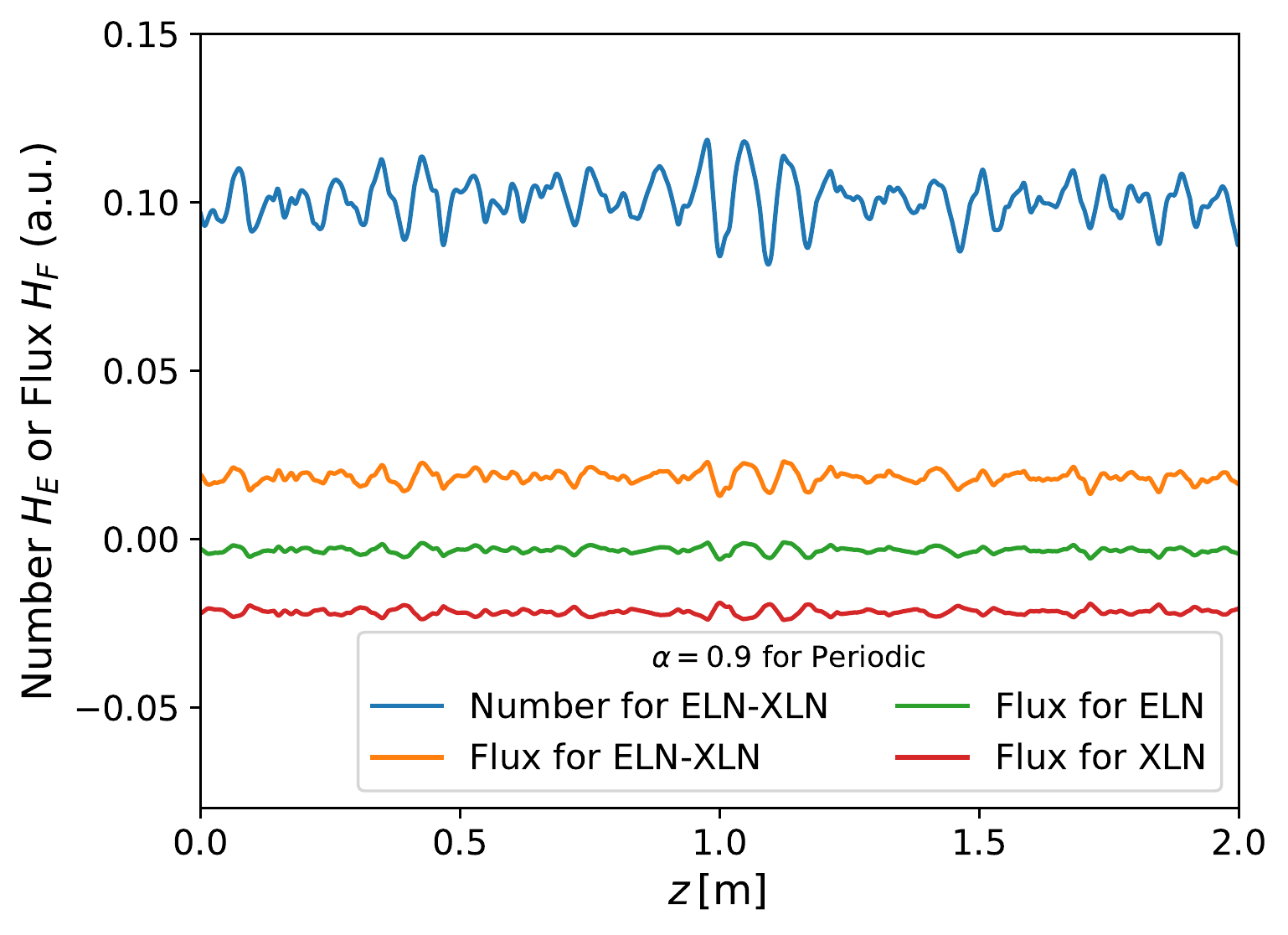}
    \includegraphics[width=0.95\linewidth]{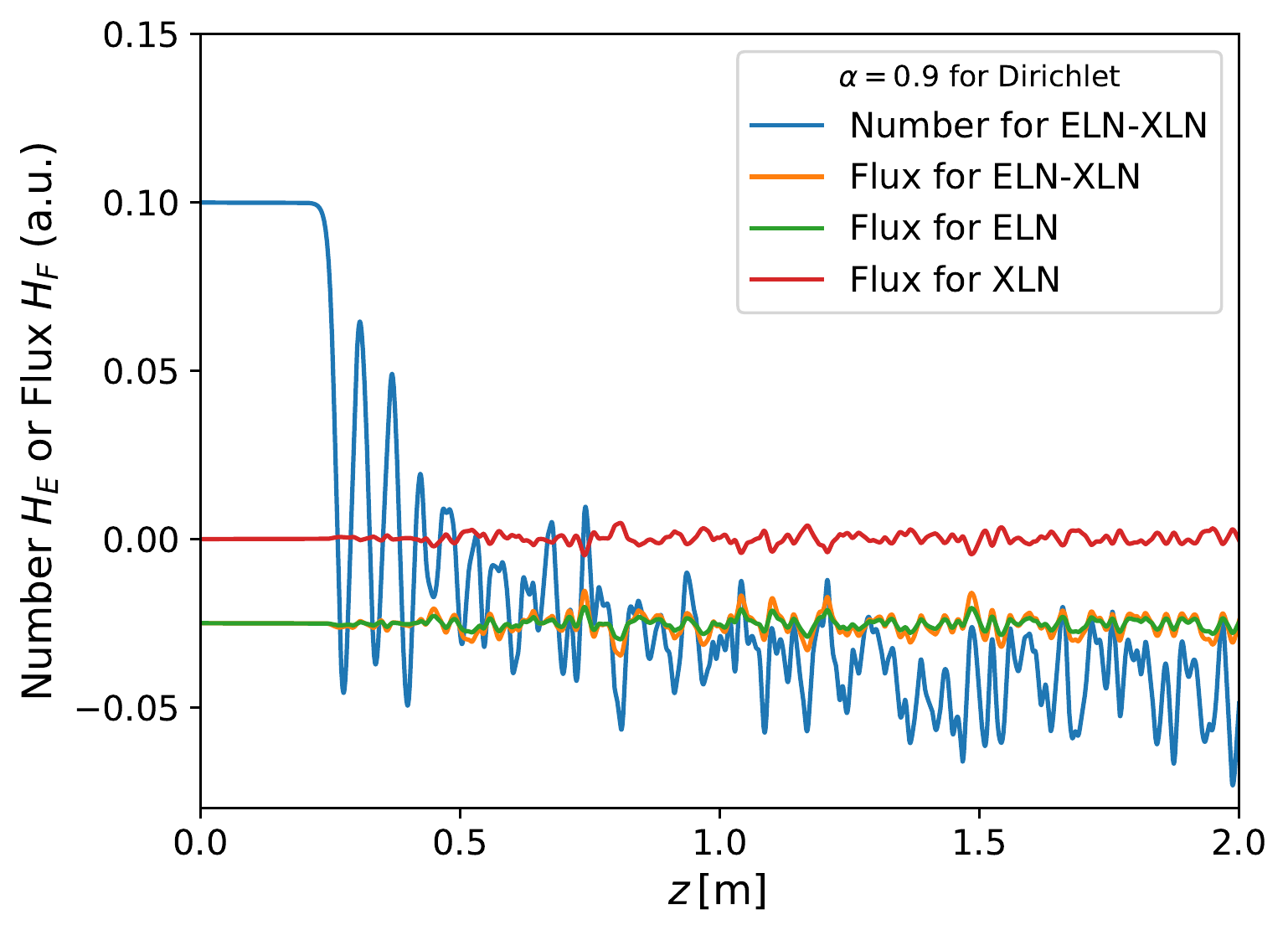}
    \includegraphics[width=0.95\linewidth]{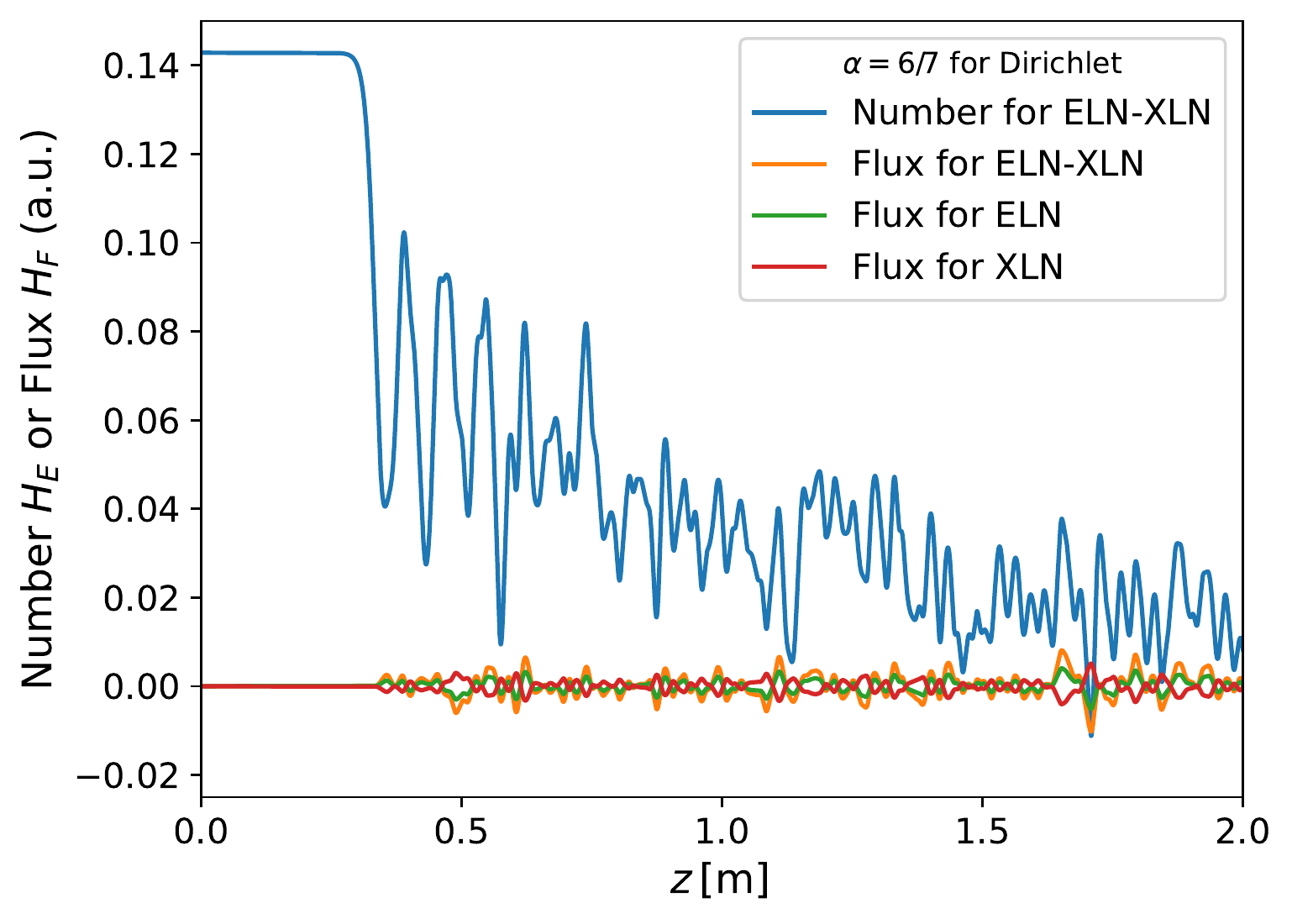}
    \caption{Spatial distributions of neutrino number density $H_E$ and number flux $H_F$ averaged during the quasi-steady state, $40\,\mathrm{ns}\leq t \leq 100\,\mathrm{ns}$, in the periodic (top) and the Dirichlet cases (middle) for $\alpha=0.9$, and in the Dirichlet case for $\alpha=6/7$ (bottom).}
    \label{fig:t_avg_flux}
\end{figure}
\begin{figure*}[t]
    \centering
    \includegraphics[width=0.45\linewidth]{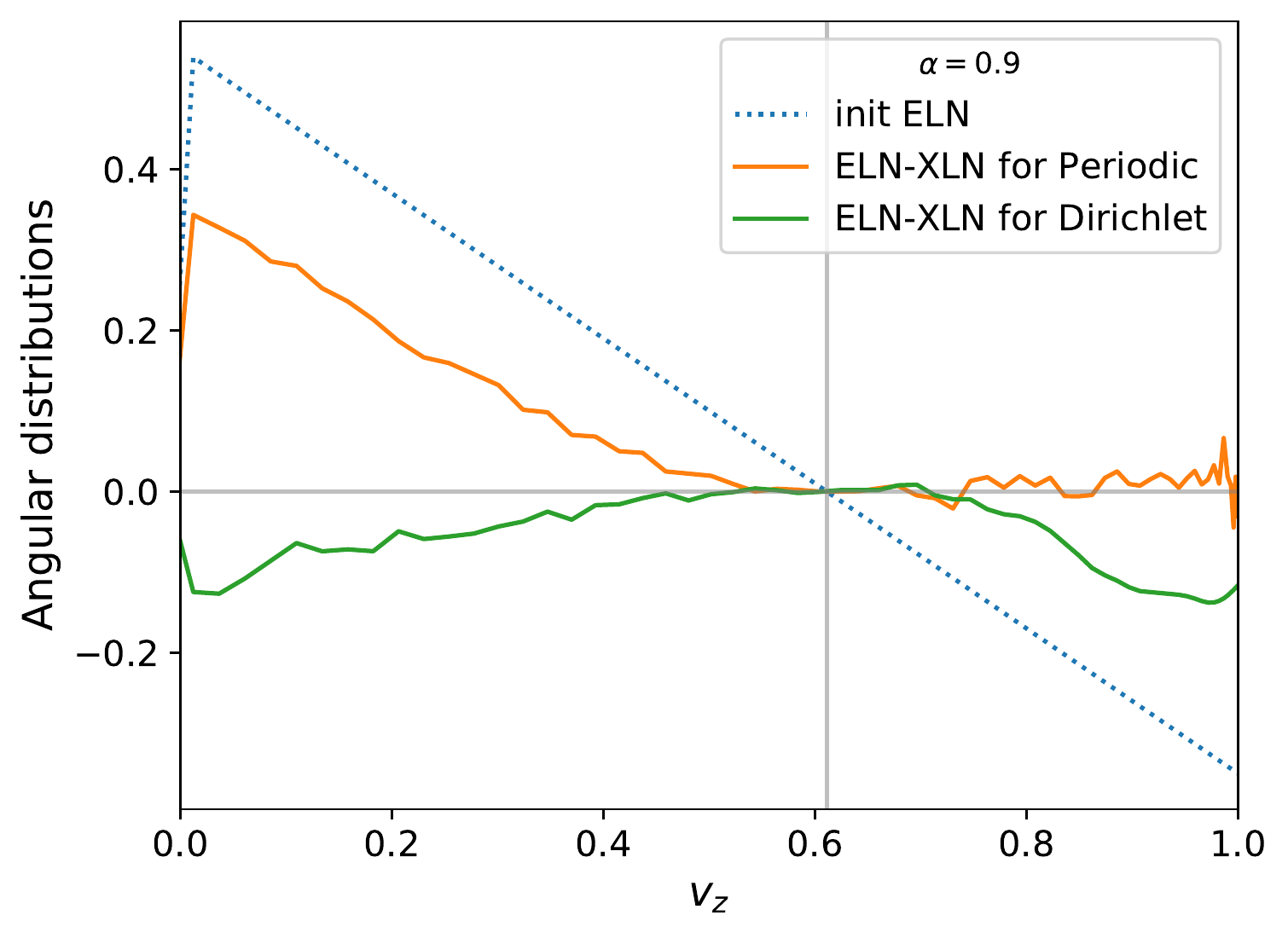}
    \includegraphics[width=0.45\linewidth]{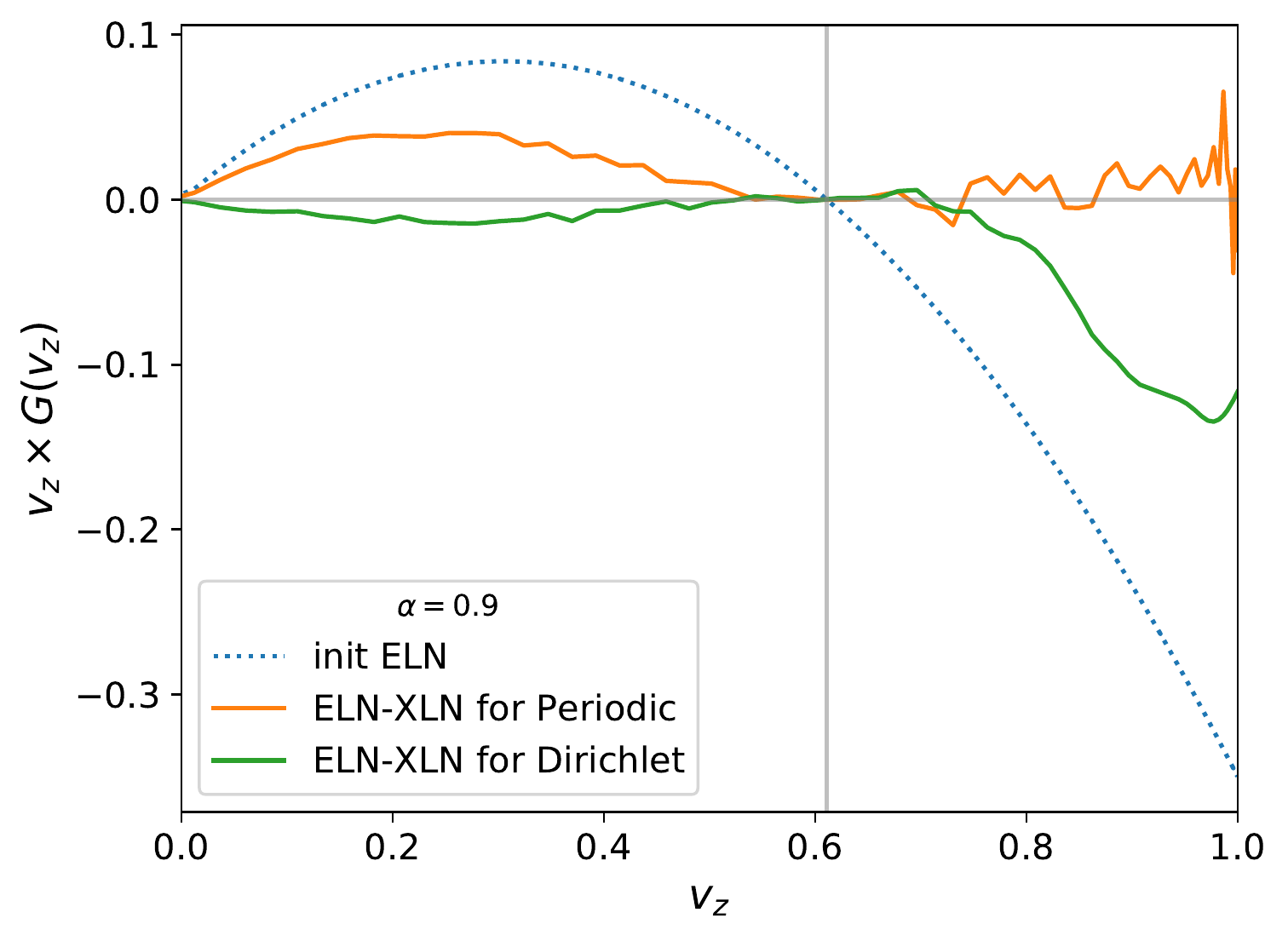}
    \includegraphics[width=0.45\linewidth]{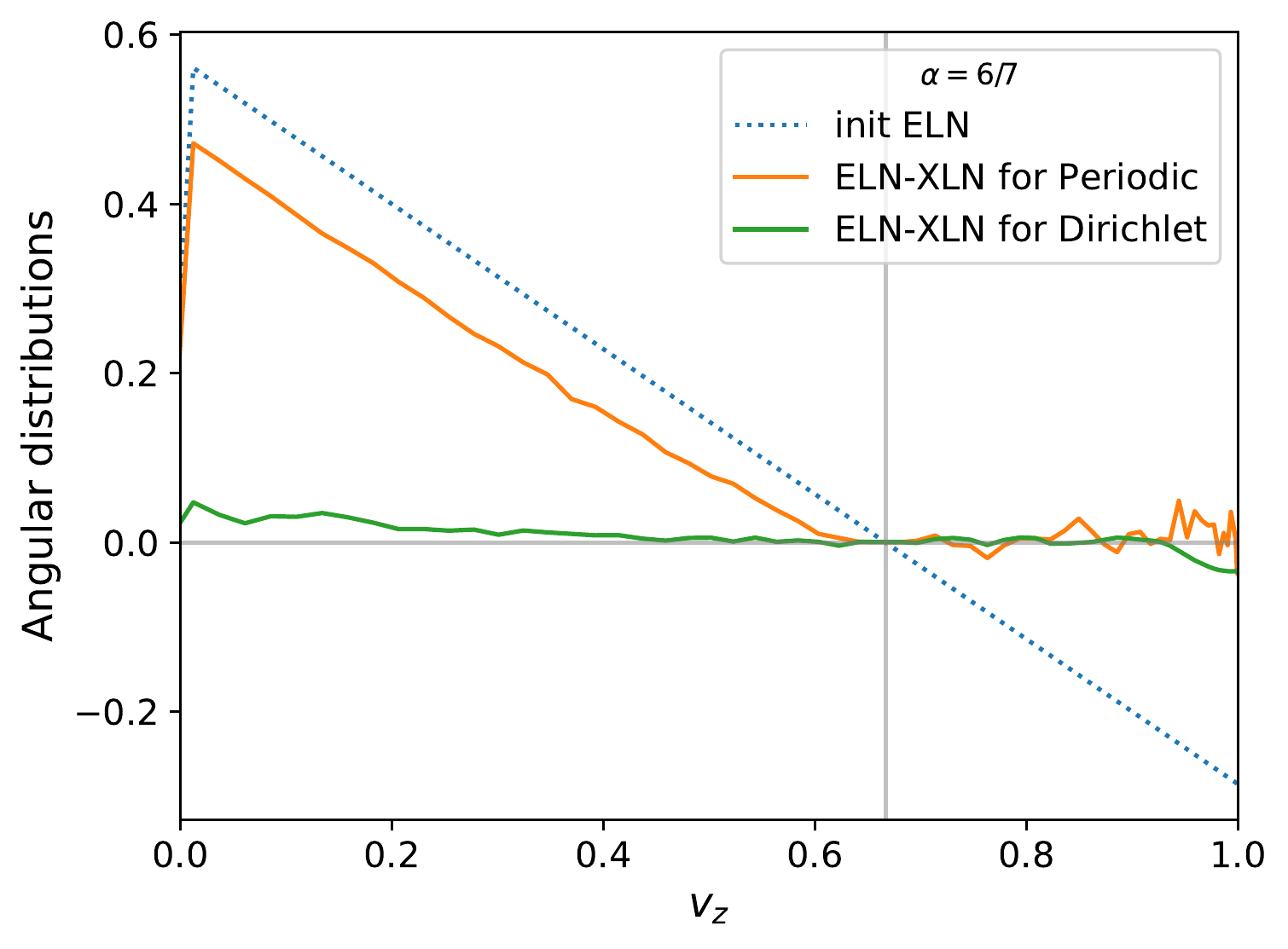}
    \includegraphics[width=0.45\linewidth]{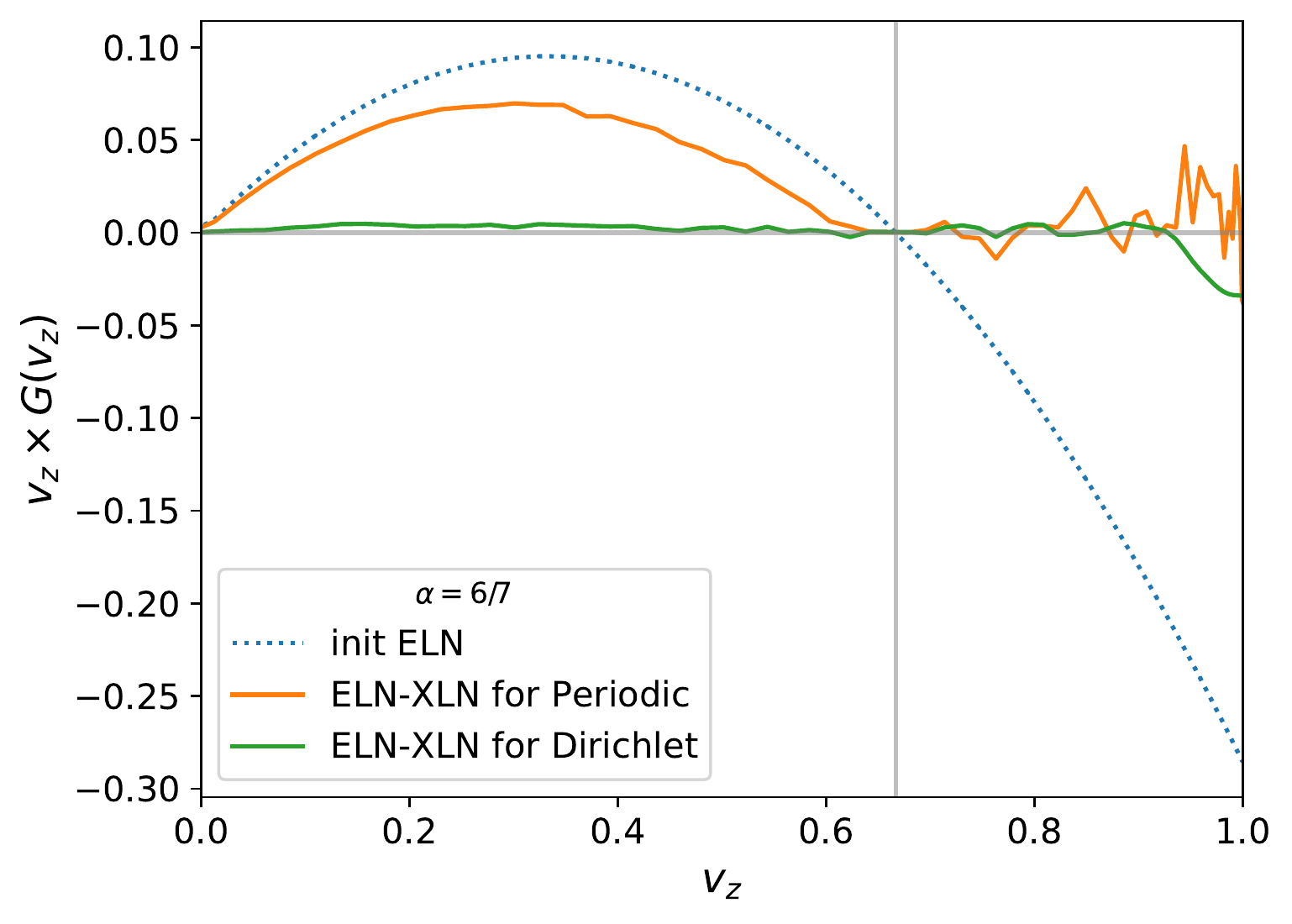}
    \caption{Time-averaged angular distributions (left panels) and ones weighted by velocity directions $v_z$ (right panels) in the window of $40\,\mathrm{ns}\leq t \leq 100\,\mathrm{ns}$ during the quasi-steady state for $\alpha=0.9$ (top) and $\alpha=6/7$ (bottom).
    Dotted line is an initial ELN angular distribution and solid green (orange) ones are time-averaged ELN-XLN angular distribution at $z=2\mathrm{~m}$ for the Dirichlet (periodic) boundary condition.}
    \label{fig:t_avg_EXLN}
\end{figure*}
\begin{figure}[t]
    \centering
    \includegraphics[width=0.95\linewidth]{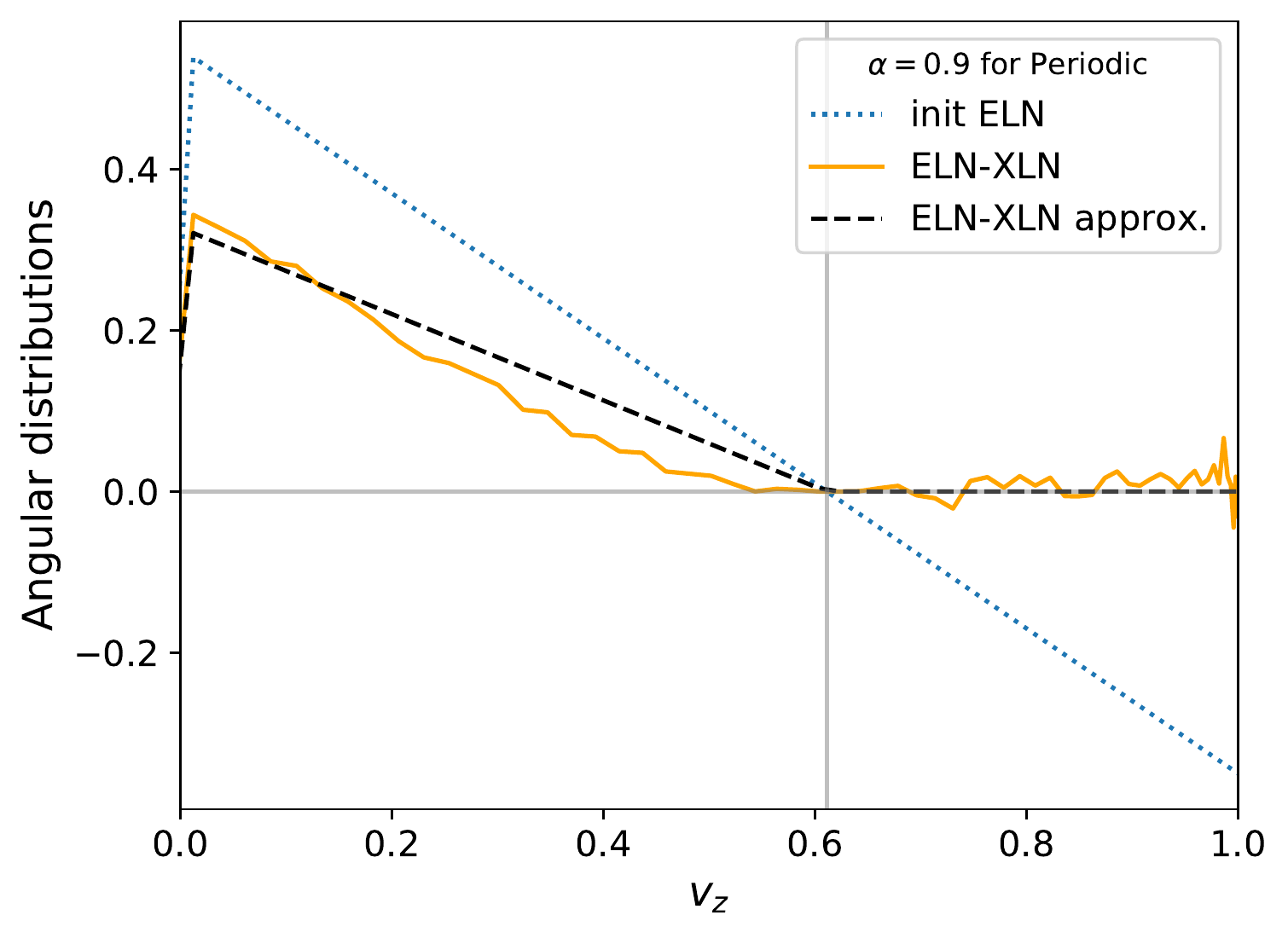}
    \includegraphics[width=0.95\linewidth]{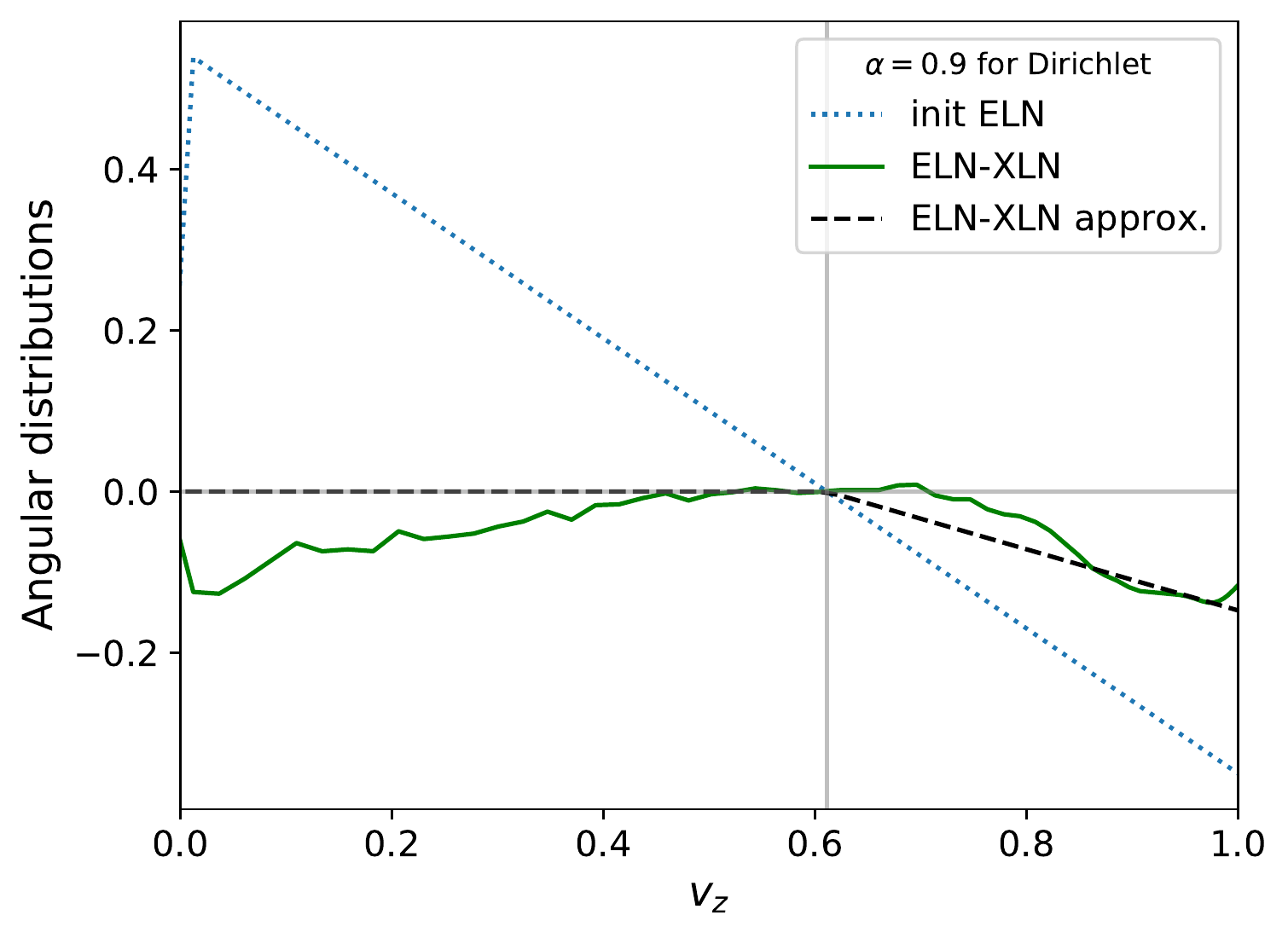}
    \caption{Time-averaged angular distributions in the window of $40\,\mathrm{ns}\leq t \leq 100\,\mathrm{ns}$ during the quasi-steady state for $\alpha=0.9$ at $z=2\mathrm{\,m}$ in the periodic (top) and Dirichlet case (bottom).
    Solid line is for numerical simulations and dashed line is for our analytical scheme.}
    \label{fig:t_approx_EXLN}
\end{figure}

\section{Numerical simulations}\label{Sec.III:NumSim}
\subsection{Setups}\label{subsec:setup}
In the following, our assertion that neutrino distributions in quasi-steady state can be determined analytically is tested through numerical simulations.
In this work, we ignore the matter term for simplicity and set a small mixing angle in the vacuum term mimicking the matter suppression instead.
Also, the entire system becomes almost energy-independent because FFC is driven only by the self-interaction potential, and so the vacuum term is considered only to trigger FFC as a perturbation.
Thereby, we set $\Delta m^2 = 2.5\times 10^{-6}\mathrm{~eV^2}$, $\theta_{\mathrm{mix}}=10^{-6}$, and $E = 12\mathrm{~MeV}$ as a monochromatic assumption.

We carry out FFC simulations for both periodic and Dirichlet boundary conditions.
In the periodic case, the numerical setup is the same as that used in our previous paper \cite{Zaizen:2023a}.
In the Dirichlet case, we consider incident neutrinos from the boundary ($z=0$) to be Dirichlet and outgoing ones to be free streamingly.
We also note that the neutrinos are dominated by $v_z > 0$; more specifically, we set dilute neutrinos in inward directions, which is the same setup used in Refs.\,\cite{Nagakura:2022a,Nagakura:2023a}.
Angular distributions for $\nu_e$ and $\bar{\nu}_e$ are given by
\begin{align}
f_{\nu_e}(v_z) &= 
\begin{cases}
    1  &\mathrm{for}\,\,v_z \geq 0 \\
    \eta &\mathrm{for}\,\,v_z < 0,
\end{cases}\\
f_{\bar{\nu}_e}(v_z) &= 
\begin{cases}
    \alpha\left(v_z + 0.5\right)  &\mathrm{for}\,\,v_z \geq 0 \\
    \alpha\eta &\mathrm{for}\,\,v_z < 0,
\end{cases}
\end{align}
where $\eta$ and $\alpha$ represent the diluteness of inward-going neutrinos and an asymmetry parameter, respectively.
We set $\eta=10^{-6}$, which is small enough that these neutrinos have negligible contributions to angular moments and neutrino self-interactions.
In our initial models, we parametrize the location of angular crossing ($v_c$) and its depth through $\alpha$.
We here employ two angular distribution models with $v_c=11/18$ for $\alpha=0.9$ and $v_c=2/3$ for $\alpha=6/7$.
We also assume no heavy-leptonic neutrinos in the initial condition.
The setup with $\alpha=6/7$ corresponds to the flux symmetric case.
According to our consideration presented in Sec.\,\ref{subsec:characterizing}, this corresponds to the case with $B_1=A_1$, in Eq.\,\eqref{eq:angular_moments}, indicating that FFC leads to a full flavor equilibrium in the entire angular region.
This emerges, indeed, in our simulation. For all simulations, we assume the number density $n_{\nu_e}=6\times 10^{32}\mathrm{\,cm^{-3}}$ in the initial condition.
We set the spatial domain of $0\leq z \leq 10\mathrm{~m}$ with a resolution $N_z=49152$ on the equal bins and angular binning $N_{v_z}=128$.

\subsection{Results}\label{subsec:results}
Before comparing to our analytic description of the quasi-steady state, we discuss some time-dependent features in the case of the Dirichlet boundary condition, which helps to deepen our understanding of how the asymptotic state is established.
Figure\,\ref{fig:t_evo_Pex} shows color maps of transition probability $P_{ex}$ for $\alpha=0.9$ (top) and $\alpha=6/7$ (bottom) as functions of $z$-axis and velocity direction $v_z$ at different time snapshots ($t=20, 60$, and $100\mathrm{\,ns}$).
The white dashed line in each panel draws the location of a crossing in the initial ELN angular distribution.
In the early epoch, flavor instabilities grow everywhere from perturbations via the vacuum term.
As shown in the left panels of Fig.\,\ref{fig:t_evo_Pex}, they occur mainly on the shallower side (in more forward directions) of the ELN angular distribution.
One thing we do notice here is that the flavor conversions temporarily establish a {\it pseudo}-asymptotic state in the region far away from the Dirichlet boundary (see the left panels in Figure\,\ref{fig:t_evo_Pex}).
This is due to the fact that the flavor conversion is so rapid to reach a nonlinear saturation by interacting with nearby flavor waves.
The resultant angular distribution is very similar to that in the case of the periodic boundary condition.
This makes sense because the flavor conversion occurs locally and the surrounding environment is also similar.
However, the flavor waves from the inner Dirichlet boundary propagate, break the periodic-like asymptotic state, and transit to another quasi-steady state.
The transition of quasi-steady state can be clearly seen in the middle and right panels in Fig.\,\ref{fig:t_evo_Pex}.

Another intriguing feature displayed in Fig.\,\ref{fig:t_evo_Pex} is that flavor waves from the Dirichlet boundary have wave-like coherent structures, particularly in $0\leq v_z < v_c$.
Small-scale structures, which appear in $v_c \leq v_z$ at the early epochs, are formed due to the interference among flavor waves in the nonlinear regime \cite{Wu:2021}.
Conversely, coherent patterns as appearing in $0\leq v_z < v_c$ imply that the flavor waves do not interact with the other components to just propagate towards the positive-$z$ direction.
In other words, the coherent flavor waves indicate that flavor conversion occurs in the angular directions opposite to the pseudo-asymptotic state in the periodic-like domain.
Note that in $v_c \leq v_z$ for $\alpha=6/7$ and around $v_c$ for $\alpha=0.9$, the interactions between flavor waves from the Dirichlet boundary and the periodic-like parts similarly generate small-scale structures.

We now move on to the in-depth analysis of the difference in the quasi-steady state between periodic and Dirichlet boundary conditions.
To this end, we focus on the time-averaged quantities during the quasi-steady state.
Figure\,\ref{fig:t_avg_Pex} shows transition probability averaged in the window of $40\,\mathrm{ns}\leq t \leq 100\,\mathrm{ns}$, in which the inner region $z\leq 2\mathrm{~m}$ has already achieved the quasi-steady state.
The top and middle panels correspond to the cases in periodic and Dirichlet boundary conditions for $\alpha=0.9$, respectively.
We note that the blue-colored region in the vicinity of the Dirichlet boundary ($z = 0$) corresponds to the region where flavor instabilities grow, and the asymptotic state in the Dirichlet case corresponds to the neutrino distribution around $z=2\mathrm{~m}$ in the figure.
By comparing between the top and middle panels, we find that the angular direction where the system achieves a flavor equipartition is opposite of each other across $\sim v_c$.
For the case of $\alpha=6/7$ with Dirichlet boundary condition, displayed in the bottom panel of Fig.\,\ref{fig:t_avg_Pex}, one can find that nearly flavor equilibrium in all angular directions is achieved at $z=2\mathrm{~m}$.
This is exactly what we expected from our analytic argument.

Figure\,\ref{fig:t_avg_flux} shows the spatial distributions of number density $H_E$ and neutrino flux $H_F$ averaged during the quasi-steady state in the periodic (top) and the Dirichlet cases (middle) for $\alpha=0.9$, and in the Dirichlet case for $\alpha=6/7$ (bottom).
As mentioned in Sec.\,\ref{Sec.II:quasi}, the flux for each flavor is almost constant in space irrespective of imposed boundary conditions.
However, comparing between the periodic and the Dirichlet cases for $\alpha=0.9$ (see top and middle panels in Fig.\,\ref{fig:t_avg_flux}), there are some notable differences.
As shown in the top panel, $H_E$ is almost constant ($\sim 0.1$) in the entire computational domain for the periodic case, while it decreases substantially at $z \sim 0.3\mathrm{\,m}$ in the case with Dirichlet one.
This is due to the fact that $H_E$ is not a conserved quantity in the latter case.
Another important difference is that neutrino fluxes are spatially constant regardless of boundary conditions but the values depend on the boundary.
In the Dirichlet boundary condition, they are the same as those set at $z=0$ (which are also the same as in the initial condition).
For instance, the flux for XLN (solid red line in the middle panel of Figure\,\ref{fig:t_avg_flux}) is zero, despite the fact that heavy leptonic neutrinos appear due to FFC.
On the other hand, the XLN flux in the periodic case (top panel) clearly deviates from zero, exhibiting that the neutrino fluxes are different from those in the initial condition.
In the case of $\alpha=6/7$ with Dirichlet boundary (bottom panel), we find that the neutrino fluxes are almost zero regardless of flavors, which is consistent with our Dirichlet boundary condition.

Figure\,\ref{fig:t_avg_EXLN} displays ELN-XLN angular distributions (left panels) and ones weighted by velocity directions $v_z$ (right panels) at $z=2\mathrm{\,m}$, in which we take a time average during the quasi-steady state, in both periodic and Dirichlet cases for $\alpha=0.9$ (top) and $\alpha=6/7$ (bottom).
We note that the angular integration of each quantity corresponds to the number and the flux for ELN-XLN, respectively, which can highlight the different properties between the periodic and Dirichlet cases.
As mentioned already, the ELN-XLN number conservation is one of the key properties to determine the quasi-steady state in the periodic case.
As shown in the orange line in the left panels of Fig.\,\ref{fig:t_avg_EXLN}, the ELN-XLN crossing disappears in the quasi-steady state while keeping the angular-integrated ELN-XLN distribution (i.e., ELN-XLN number) to be the same as the initial one (dotted line).
Indeed, the ELN-XLN in the angular region of $v_z < v_c$ in the asymptotic state needs to be smaller than the initial one because ELN-XLN in $v_z > v_c$ needs to be increased so as to erase the ELN-XLN crossing.
On the other hand, we observe that the angular-integrated ELN-XLN distributions in the Dirichlet boundary become negative and nearly zero for $\alpha=0.9$ and $6/7$, respectively, in the quasi-steady state, exhibiting that it is different from the initial one (which is positive).

On the other hand, $v_z$-weighted ELN-XLN angular distribution has the opposite trend from the above argument.
In the case of the Dirichlet boundary (green line in the right panel of Fig.\,\ref{fig:t_avg_EXLN}), its angular-integrated value (i.e., flux) is the same as the initial one, which is negative and almost zero in the case with $\alpha=0.9$ and $6/7$, respectively.
As displayed in the right panels, the ELN-XLN becomes zero in the angular region of $v_z < v_c$, and therefore the $v_z$-weighted ELN-XLN angular distribution in $v_z > v_c$ approaches to zero so as to keep the flux being constant.
For the periodic boundary condition, however, the flux becomes positive in the asymptotic state (see orange lines), exhibiting that it is not conserved.

One thing we do notice along the above argument is that if the conserved quantity (either number or flux, depending on the boundary condition) is initially zero, a flavor equipartition is established in the whole angular direction.
We also confirm that the disappearance of ELN-XLN angular crossings is a general ingredient that characterizes the quasi-steady state irrespective of imposed boundary conditions.
Satisfying the stability condition by damping the ELN-XLN angular crossing, we conclude that the conservation law determines in angular directions which a flavor equipartition is achieved.

Finally, we test our analytic model Eqs.\,\eqref{eq:approx_1} and \eqref{eq:approx_2} for both the periodic and Dirichlet boundaries in the case with $\alpha=0.9$;
the result is displayed in Fig.\,\ref{fig:t_approx_EXLN}.
We find that our model (black dashed line) can capture the characteristics in the quasi-steady state of FFC for both boundary conditions.
Note that some minor deviations in $0\leq v_z < v_c$ from numerical results come from the spatial variations as noted for Fig.\,\ref{fig:t_avg_Pex}.
This again indicates that different boundary conditions differ only in what is conserved in the simulation box, and that the mechanism leading to the quasi-steady state is essentially the same: the stability and the conservation laws.

\section{Conclusions}\label{sec:conclusion}
In this paper, we generalize our previous study to determine the quasi-steady states of fast neutrino-flavor conversion (FFC) based on the argument with stability and conservation laws.
This argument reveals the physical reason for how the boundary condition affects the quasi-steady state.
This is due to the difference in conserved quantity: 
ELN-XLN number and flux for periodic and Dirichlet boundary conditions, respectively.
Based on the analytic argument, we provide a simple analytic scheme to determine the asymptotic state of FFC, which can be applied to both the periodic- and Dirichlet boundary cases.
By performing numerical simulations of FFC in a local one-dimensional box, we show that our analytic model has the ability to capture the essential trend of the asymptotic state.

Our time-dependent numerical simulations also exhibit how the system approaches a quasi-steady state, depending on the boundary condition.
In the periodic one, FFC experiences a nonlinear saturation due to a cascade through interactions among nearby flavor waves and works to eliminate ELN-XLN angular crossings to achieve the stability.
Then, the number conservation for both ELN and XLN leads to a flavor equipartition on the shallower side of ELN-XLN angular distribution.
In the case of the Dirichlet boundary condition, we find an intriguing temporal feature before reaching the final asymptotic state.
At the early epoch, the system locally mimics the periodic cases and temporarily reaches a pseudo-asymptotic state following the number conservation.
However, the quasi-steady state is gradually repainted over time after the neutrinos injected with the Dirichlet boundary condition reach there.
We demonstrate such a transition from the pseudo-quasi-steady state to the final one, in which the angular distributions of neutrinos become very distinct from each other.

This study suggests that if the sign of the flux ($H_F$) is opposite to that of the number ($H_E$), the angular direction where flavor equipartition occurs should be opposite of each other.
Also, when the initial flux is zero, the entire system establishes a full flavor equilibrium in the case of the Dirichlet boundary.
This mechanism is essentially the same as in the periodic case for $\alpha=1$, with symmetry between neutrinos and antineutrinos.
The only difference is whether they impose a conservation law on the number or the flux, while the disappearance of ELN-XLN angular crossings remains a general condition determining the quasi-steady state.
Note that the flux becomes spatially constant in the quasi-steady state even under the periodic boundary condition.
However, unlike the Dirichlet case, neutrinos distributions are not fixed at the boundaries, and the flux is not guaranteed to be identical to that in the initial condition.
Thereby, the flux is spatially constant but not a temporally-conserved quantity in the periodic boundary condition.

Although we provide important clues in understanding the asymptotic states of FFC, several improvements remain left behind.
One of the shortcomings in this study is that we neglect the collision term.
We also simplify the system by employing the dilute inward-going neutrinos and the one-dimensional spatial coordinate.
Considering more general cases requires some improvements in both our approximate scheme and numerical simulations.
We leave the detailed study to future work, which will be reported in our forthcoming paper(s).

\begin{acknowledgments}
We thank the Focus workshop on collective oscillations and chiral transport of neutrinos at the Academia Sinica for eliciting the discussions that led to the completion of this paper.
We are grateful to Huaiyu Duan, Meng-Ru Wu, Lucas Johns, Sajad Abbar, and Chinami Kato for useful comments and discussions.
M.Z. is supported by the Japan Society for Promotion of Science (JSPS) Grant-in-Aid for JSPS Fellows (Grants No. 22J00440) from the Ministry of Education, Culture, Sports, Science, and Technology (MEXT) in Japan.
The numerical computations were carried out on Cray XC50 at the Center for Computational Astrophysics, National Astronomical Observatory of Japan.
H.N is supported by Grant-in-Aid for Scientific Research (23K03468) and by the HPCI System Research Project (Project ID: 220173, 220047, 220223, 230033).
\end{acknowledgments}


\bibliographystyle{apsrev4-1}
\bibliography{boundary}

\end{document}